\documentclass[aps,10pt,prb,twocolumn,groupedaddress,showpacs]{revtex4}

\usepackage{graphicx}
\usepackage[caption=false]{subfig}
\usepackage{amsmath}
\usepackage{bbold}
\usepackage{latexsym}

\begin{document}

%\preprint{}

\title{Probing the $\nu=\frac{5}{2}$ quantum Hall state with electronic Mach-Zehnder interferometry}


\author{Guang Yang}
%\email[]{}
%\thanks{}
\affiliation{RIKEN Center for Emergent Matter Science, Wako, Saitama 351-0198, Japan}


\date{\today}

\begin{abstract}

It was recently pointed out that Halperin's 113 topological order explains the transport experiments in the quantum Hall liquid at filling factor $\nu=5/2$. The 113 order, however, cannot be easily distinguished from other likely topological orders at $\nu=5/2$ such as the non-Abelian Pfaffian and anti-Pfaffian states and the Abelian Halperin 331 state in Fabry-Perot interferometry. We show that an electronic Mach-Zehnder interferometer provides a clear identification of these candidate $\nu=5/2$ states. Specifically, the $I$-$V$ curve for the tunneling current through the interferometer is more asymmetric in the 113 state than in other $\nu=5/2$ states. Moreover, the Fano factor for the shot noise in the interferometer can reach 13.6 in the 113 state, much greater than the maximum Fano factors of 3.2 in the Pfaffian and anti-Pfaffian states and 2.3 in the 331 state.

\end{abstract}


% insert suggested PACS numbers in braces on next line \pacs{}
%\keywords{}

\maketitle

% References should be done using the \cite, \ref, and \label command

\section{Introduction}

The fractional quantum Hall (FQH) state at filling factor $\nu=5/2$ has attracted much interest since its discovery \cite{willett87}. Unlike the common FQH states at filling factors with odd denominators, the $\nu=5/2$ FQH state cannot be explained by a hierarchical construction \cite{wenbook} of variational wave functions based on the Laughlin state. The fact that the filling factor has an even denominator indicates the possibility of electron pairing. Along this line, a number of models \cite{moore91,jain89,wen91,levin07,lee07,yang13,halperin83,wen92,overbosch} were proposed to explain the $\nu=5/2$ FQH state (see Ref.~\onlinecite{yang13} for an overview of the proposed models). In some of those proposals, quasiparticle excitations with non-Abelian statistics were predicted. A collection of non-Abelian quasiparticles span a degenerate ground-state manifold which may be useful for topological quantum computation \cite{kitaev03,nayak08}. Such non-Abelian models \cite{moore91,jain89,wen91,levin07,lee07,yang13} include the Pfaffian state, the $SU(2)_2$ state, the anti-Pfaffian state, and the anti-$SU(2)_2$ state. At the same time, models \cite{halperin83,wen92,yang13} predicting ``ordinary'' Abelian quasiparticles, such as the Halperin 331 state, the $K=8$ state, and their particle-hole dual states, were also constructed.

In all the proposed models, a fundamental quasiparticle charge of $e/4$ was predicted. This fundamental charge follows from a general argument \cite{levin09} for the even-denominator quantum Hall states and has been observed experimentally \cite{dolev08,willett09,lin12}. On the other hand, different models have different implications for the topological nature of the $\nu=5/2$ state. Several experiments \cite{lin12,radu08,baer14,bid10,mstern10,rhone11,tiemann12,mstern12,chickering13,willett10} were designed to probe the topological order at $\nu=5/2$.
References~\onlinecite{lin12,radu08,baer14} measured the temperature and voltage dependence of quasiparticle tunneling through a quantum point contact. In the weak-tunneling regime, the zero-bias tunneling conductance $G$ scales with temperature $T$ according to a power law, $G\sim T^{2g-2}$, where the exponent $g$ depends on the topological order in the bulk, a manifestation of the edge-bulk correspondence \cite{wenbook} in FQH systems. The results of the tunneling experiments were argued \cite{yang13} to agree with the Halperin 331 state after taking into account the effect of long-range electrostatic interaction near the tunneling point. On the other hand, the chiral 331 state does not explain the observation \cite{bid10} of an upstream neutral mode on the edge of the $\nu=5/2$ liquid. Overall, none of the above-mentioned models seems to fit in the constraints set by the experiments, assuming the effect of edge reconstruction is less important. Edge reconstruction is likely in a pure $\nu=5/2$ liquid \cite{addwan06,addzhang14} but is expected to be suppressed in real samples with disorder, as is evidenced by the recent experiment \cite{addinoue14} showing relatively weak signals associated with edge reconstruction and that such signals disappear at long distances \cite{bid10}.

In a recent paper \cite{yang14}, it was argued that Halperin's 113 topological order  provides a consistent explanation of the transport data in the $\nu=5/2$ FQH liquid. The 113 order is Abelian and comes with both spin-unpolarized and spin-polarized versions. It supports an upstream neutral edge mode and predicts the correct scaling behavior of the zero-bias conductance observed in the tunneling experiments. Distinguishing the 113 state from the other possibilities, especially the 331 state, is a subtle experimental task. The predictions of the 113 state and the 331 state are quite close in the tunneling experiments, whose difference lies within experimental uncertainty. \cite{yang13,yang14}  The measurements of spin polarization \cite{mstern10,rhone11,tiemann12,mstern12} are controversial and do not help, since the 113 state and the 331 state allow both zero- and full- spin polarizations \cite{yang13,overbosch,yang14}. Data of bulk thermopower measurement \cite{chickering13} showed features that may be associated with non-Abelian quasiparticles, \cite{addyang09} but the 113 and 331 states may exhibit similar features if they host different quasiparticle species that are degenerate in energy. Results \cite{willett09,willett10} of an electronic Fabry-Perot interferometer were interpreted \cite{bishara09} to be compatible with the non-Abelian states. However, the 113 state and the 331 state can produce similar signatures, in the presence of approximate or exact symmetry in the tunneling behaviors of different species of quasiparticles. \cite{astern10,yang14} The existence of an upstream neutral edge mode favors the 113 state and the anti-Pfaffian state over the 331 state and the Pfaffian state. However, more experimental effort based on a variety of methods \cite{up1,up2,up3,fdt1,fdt2,fdt3} is needed before one can draw a definite conclusion about the existence of an upstream mode. Thus, it is necessary to have an alternative approach which offers additional data to test the proposal of the 113 topological order.

In this paper, we show that an electronic Mach-Zehnder interferometer \cite{ji03,neder07,weisz14,feldman14,law06,feldman06,feldman07,law08,wang10,campagnano12,ponomarenko07,ponomarenko10} provides a clear identification of the 113 state, the 331 state, and the Pfaffian state, while it exhibits identical characteristics in the Pfaffian and anti-Pfaffian states. We have not included other $\nu=5/2$ states in the analysis, because they are less probable candidates as revealed by the experiments \cite{lin12,radu08,baer14,bid10}. We compute the tunneling current through the interferometer in the 113 state and compare it with the currents \cite{feldman06,wang10} in the 331 state  and in the Pfaffian (or anti-Pfaffian) state. In all of the four topological orders considered, the current depends periodically on the magnetic flux enclosed by the interferometer and is asymmetric under the change of the sign of the applied voltage. The $I$-$V$ curve is most asymmetric in the 113 state. We have also studied the low-frequency shot noise of the tunneling current and found that the Fano factor, defined as the noise-to-current ratio, also oscillates periodically with the magnetic flux. The Fano factor can achieve 13.6 in the 113 state, much greater than the maximum achievable Fano factors in the 331 state and in the Pfaffian (or anti-Pfaffian) state, which were previously found \cite{feldman07,wang10} to be 2.3 and 3.2, respectively. These results, based on quasiparticle braiding statistics, are not sensitive to the edge structure, including edge reconstruction. Thus, a Mach-Zehnder interferometer can serve as a useful probe of the topological order at $\nu=5/2$. 

The paper is organized as follows. In Sec.~II, we review the statistical properties of the 113 topological order. In Sec.~III, we explain the structure of an electronic Mach-Zehnder interferometer and its operation in the 113 state. In Secs.~IV and V, we calculate the tunneling current and shot noise in the 113 state at zero temperature, and compare the results with those obtained in the 331 state and in the Pfaffian and anti-Pfaffian states. The zero-temperature limit describes the situation where the temperature is much lower than the applied voltage in the interferometer. We explain the reason for the asymmetric $I$-$V$ curve and large shot noise in the 113 state. We  conclude in Sec.~VI.

\section{Statistical properties of the 113 topological order}
The 113 topological order has a spin-unpolarized version and a spin-polarized version. \cite{yang14} Its statistical properties are captured by the standard K-matrix formalism \cite{wenbook}.

 The spin-unpolarized 113 state has the K matrix
\begin{equation}
\label{y1}
K =
\left( \begin{array}{cc}
1 & 3 \\
3 & 1 \end{array} \right),
\end{equation}   
which encodes information about quasiparticle statistics, and the charge vector $\mathbf{q}=(1,1)$, which describes how the excitations couple to the electromagnetic gauge field. Its two fundamental quasiparticles, denoted by $a$ and $b$, are represented by the vectors $\mathbf{l}_{a}=(1,0)$ and $\mathbf{l}_{b}=(0,1)$, respectively, both with a quarter electron charge. The statistical phase a fundamental quasiparticle acquires after it makes a full circle around another fundamental quasiparticle of different or the same flavor is  
\begin{equation}
\label{y2}
\phi_{ab}=\phi_{ba}=2\pi \mathbf{l}_a K^{-1}  \mathbf{l}_b^T=3\pi/4
\end{equation}
or 
\begin{equation}
\label{y3}
\phi_{aa}=\phi_{bb}=2\pi \mathbf{l}_{a} K^{-1}  \mathbf{l}_{a}^T=-\pi/4,
\end{equation}
respectively.
The two flavors of the fundamental quasiparticles in the spin-unpolarized 113 state may be intuitively understood as being related to the two electron spin species. A generic quasiparticle excitation can be viewed as a linear combination of the fundamental quasiparticles. For instance, the electron operators, defined as quasiparticles having unit electron charge and obeying fermionic statistics, are represented by the vectors $(3,1)$ and $(1,3)$.

The spin-polarized 113 state can be interpreted as a hierarchical FQH state, formed by condensing charge-$2e$ quasiholes on top of a $\nu=1$ integer quantum Hall state. Its K matrix is
\begin{equation}
\label{y4}
K' =
\left( \begin{array}{cc}
1 & 2 \\
2 & -4 \end{array} \right)
\end{equation}   
and charge vector is $\mathbf{q}'=(1,0)$. The two fundamental quasiparticles in the spin-polarized 113 state are represented by the vectors $(0,1)$ and $(1,-1)$, both with $e/4$ charge and the same statistical phases as described in Eqs.~(\ref{y2}) and (\ref{y3}).

The two 113 states belong to the same topological order. \cite{yang14} Indeed, their K matrices and charge vectors are related by an $SL(2,\mathbb{Z})$ transformation $W$: $K=W^TK'W$ and $\mathbf{q}=\mathbf{q}'W$, where 
\begin{equation}
\label{y5}
 W =
\left( \begin{array}{cc}
1 & 1 \\
0 & 1 \end{array} \right).
\end{equation} 
This means that the two states have the same collection of quasiparticle species in terms of charge and braiding statistics. As a result, Mach-Zehnder interferometry based on quasiparticle braiding is unable to distinguish between the spin-unpolarized and spin-polarized 113 states. In the following sections, we discuss the tunneling current and shot noise in the context of the spin-unpolarized 113 state.

\section{Electronic Mach-Zehnder interferometer}

\begin{figure}
\centering
\includegraphics[width=2.8in]{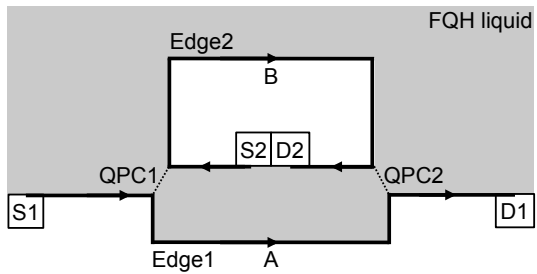}
\caption{The structure of an electronic Mach-Zehnder interferometer. The arrows denote the direction of charge propagation from the sources S1 and S2 to the drains D1 and D2. Quasiparticles can tunnel at the quantum point contacts QPC1 and QPC2, between Edge1 and Edge2. }
\end{figure}

The structure of a Mach-Zehnder interferometer is sketched in Fig.~1. Charge propagates from source S1 to drain D1 and from source S2 to drain D2 along the FQH edges Edge1 and Edge2, respectively, as indicated by the arrows. In Fig.~1, $A$ and $B$ are two points on Edge1 and Edge2, respectively. Quasiparticles can tunnel at the quantum point contacts QPC1 and QPC2. In the 113 state, the most relevant quasiparticles that participate in tunneling at low temperatures are the fundamental quasiparticles $a,b$ with $e/4$ charge. \cite{yang14} In a FQH liquid, low-energy excitations only exist on the edge. The Hamiltonian of the interferometer has the form 
\begin{equation}
\label{y6}
\hat{H}=\hat{H}_{\textrm{edge}}+\sum_{\mu=a,b}[(\Gamma_1^{\mu} \hat{T}_1^{\mu}+\Gamma_2^{\mu} \hat{T}_2^{\mu})e^{-i\frac{eV}{4} t}+\textrm{H.c.}],
\end{equation}
where $\hat{H}_{\textrm{edge}}$ is the Hamiltonian of Edge1 and Edge2, $\hat{T}_{1,2}^{\mu}$ are the tunneling operators for quasiparticle flavor $\mu$ at QPC1 and QPC2 with tunneling amplitudes $\Gamma_{1,2}^{\mu}$, and $eV$ is the chemical potential difference between the two FQH edges. If S1 is kept at a higher chemical potential than S2, then there is a net flow of quasiparticles from Edge1 to Edge2, eventually absorbed by D2. In writing the Hamiltonian we have set $\hbar=1$.

The tunneling amplitudes for different quasiparticle flavors are in general different. An interesting situation happens when both flavors of fundamental quasiparticles have identical tunneling behaviors, $\Gamma_{1,2}^a=\Gamma_{1,2}^b$. In such a limit, the interferometer exhibits elegant transport functions, as we show in Secs. IV and V. 

We assume small tunneling amplitudes at the point contacts so that the quasiparticle tunneling rate between Edge1 and Edge2 can be calculated using perturbation theory. The assumption means that the average time between two consecutive tunneling events at the point contacts is much longer than the duration of an individual tunneling event. Moreover, we assume that the tunneled quasiparticles are fully absorbed by the drain D2, leaving only their topological charges, characterized by their statistical phases. With this assumption, individual tunneling events can be considered independent. The residual topological charge at D2 can be understood as a result of the entanglement between Edge1 and Edge2: The topological charges on the two edges must add up to vacuum. The quasiparticle tunneling rate depends on the Aharonov-Bohm (AB) flux $\Phi$ enclosed by the loop $A$-QPC2-$B$-QPC1-$A$ in the interferometer, the topological charge accumulated at D2 after the previous tunneling event and the flavor of the quasiparticle being tunneled. For a quasiparticle of type $\mu$, the tunneling rate from Edge1 to Edge2 is found to be
\begin{equation}
\label{y7}
w^+_{s\rightarrow s+\mu}=\alpha(|\Gamma_1^{\mu}|^2+|\Gamma_2^{\mu}|^2)+(\beta \Gamma_1^{\mu *}\Gamma_2^{\mu } e^{i (\phi_{\textrm{AB}}+\phi_{\mu s})}+\textrm{c.c.}),
\end{equation}
where $s$ and $s+\mu$ are the topological charges at D2 before and after the tunneling event, respectively; $\phi_{\mu s}$ is the statistical phase acquired by quasiparticle $\mu$ after it makes a full circle about the topological charge $s$ at D2; $\phi_{\textrm{AB}}=2\pi\Phi/(4\Phi_0)$ with $\Phi_0=hc/e$ the flux quantum is the Aharonov-Bohm phase due to the magnetic flux through the interferometer; and $\alpha,\beta$ are functions of the voltage bias $V$, the temperature $T$, and the form factor of the interferometer, assumed independent of the quasiparticle flavor $\mu$ for simplicity. For our purpose, we do not need the explicit expressions of $\alpha,\beta$, which depend on the details in the Hamiltonian (cf. Ref.~\onlinecite{law06}). However, we point out that $\alpha,\beta$ are in principle not sensitive to the absolute distances between QPC1 and QPC2, but depend on the difference of the distances between the QPCs along different FQH edges. This property is a general advantage of Mach-Zehnder interferometry, \cite{law06,marquardt04,forster05} which allows for the observation of quantum interference at large system sizes.  From Eq.~(\ref{y7}), we see that the tunneling rate depends on the history of quasiparticle tunneling through the interferometer.

At finite temperature, quasiparticle tunneling happens from Edge2 at a lower chemical potential to Edge1 at a higher chemical potential. The tunneling rate for such an inverse tunneling process is related to that for tunneling from Edge1 to Edge2 by the principle of detailed balance: $w^-_{s+\mu\rightarrow s}=\exp[-eV/(4 k_BT)] w^+_{s\rightarrow s+\mu}$. At low temperatures, $w^-_{s+\mu\rightarrow s}$ is suppressed. We assume in the later calculations that the temperature is much lower than the applied voltage at the quantum point contacts, so that $w^-_{s+\mu\rightarrow s}$ can be neglected.

In Sec.~V, we study the shot noise of the tunneling current through the interferometer. We focus on the noise at low frequency. As was shown in Ref.~\onlinecite{feldman07}, high-frequency noise does not carry information about quasiparticle statistics, while it manifests the fractional charge of the tunneled quasiparticles.

To calculate tunneling current and shot noise in the Mach-Zehnder interferometer, one needs to understand the topological degeneracy as seen by the tunneling quasiparticle, i.e., all possible inequivalent topological charges that can be present at drain D2. In the 113 state, these topological charges are linear combinations of the fundamental quasiparticles. Assuming there have been $N_a$  quasiparticles of type $a$ and $N_b$ quasiparticles of type $b$ absorbed by D2, their total topological charge can be represented by $[N_a,N_b]$. Certain linear combinations result in trivial topological charge (trivial statistical phase as the tunneling quasiparticle encircles D2), for instance, $[3,1]$, $[1,3]$, and their integer multiples. The inequivalent topological charges are defined as $[N_a,N_b] \pmod{[3,1],[1,3]}$. In Abelian states, the fusion channels of quasiparticles are unique and the topological degeneracy admits the algebraic structure of a finite Abelian group, encoded in the K matrix. The level of degeneracy, or the group order, equals the determinant of the K matrix, \cite{wen90} given the topology of the interferometer in Fig.~1. For the 113 state, the group is $\mathbb{Z}_8$ with the generator being either of the fundamental quasiparticles. Tunneling of quasiparticles defines multiplication of group elements. In Fig.~2(a), we show the structure of topological degeneracy in the 113 state. We use solid arrows and dashed arrows to denote the transitions between inequivalent states due to tunneling of quasiparticles $a$ and $b$, respectively, at zero temperature. The tunneling current and shot noise measured at D2 are the averaged quantities over all inequivalent states in the degeneracy. 

The classification of the algebraic structure of topological degeneracy does not alone determine the current and noise in the interferometer. One also needs the explicit transition rates between inequivalent states in the degeneracy, defined by the quasiparticle tunneling rates at the point contacts. In Table~1, we list the transition rates at zero temperature. We define a set of functions
\begin{equation}
\label{y8}
p_k^{\mu}=R^{\mu}[1+c^{\mu}\cos(2\pi\Phi/(4\Phi_0)+\pi k/4+\delta^{\mu})],
\end{equation}
where $k=0,1,\dots, 7$, $R^{\mu}=\alpha(|\Gamma_1^{\mu}|^2+|\Gamma_2^{\mu}|^2)$, $c^{\mu}= 2 |\beta \Gamma_1^{\mu}\Gamma_2^{\mu }|/[\alpha(|\Gamma_1^{\mu}|^2+|\Gamma_2^{\mu}|^2)]$, and $ \delta^{\mu}=\arg(\beta\Gamma_1^{\mu *}\Gamma_2^{\mu } )$. In the presence of flavor symmetry in quasiparticle tunneling, $\Gamma_{1,2}^a=\Gamma_{1,2}^b$, we define $R=R^a=R^b$, $c=c^a=c^b$, $\delta=\delta^a=\delta^b$, and $p_k=p_k^a=p_k^b$, and draw the kinetic diagram in Fig.~2(b), where we have labeled explicitly the transition rates. We merged the vertices representing topological charges $[1,0]$ and $[0,1]$ into a single vertex, because the two vertices are identical from a kinetics point of view. The same happened to the vertices $[3,0]$ and $[2,1]$. We emphasize that there are always eight inequivalent states in the topological degeneracy, whether or not there is flavor symmetry.

\begin{figure}
\centering
\subfloat[]{\includegraphics[width=1.7in]{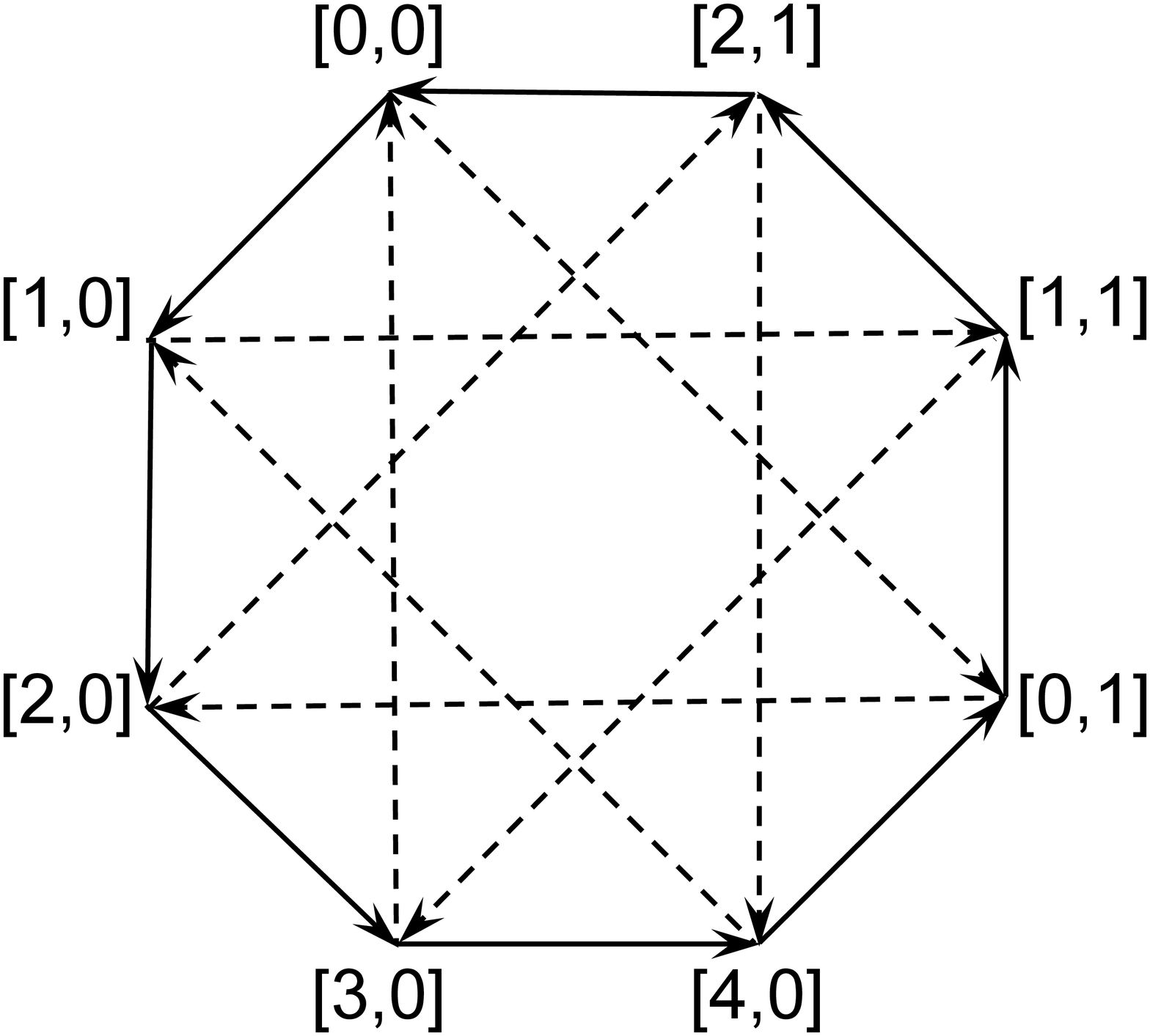}}
\quad
\subfloat[]{\includegraphics[width=1.3in]{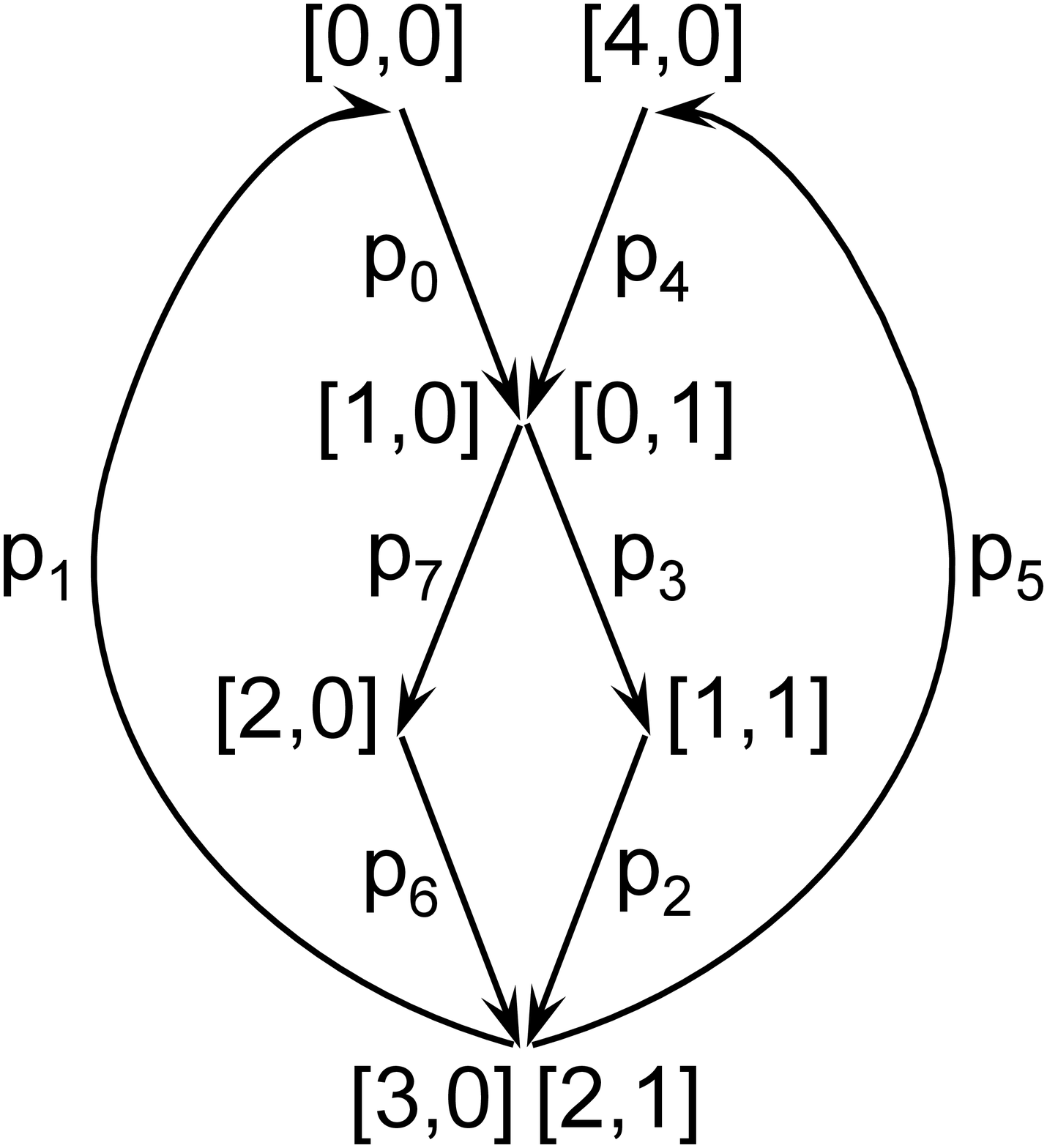}}
\caption{The structure of topological degeneracy in the 113 state (a) in the general situation and (b) in the presence of flavor symmetry. The vertices represent the eight inequivalent states in the degeneracy. The solid and dashed arrows denote transitions due to tunneling of quasiparticles $a$ and $b$, respectively, at zero temperature. Transition rates are labeled explicitly for the flavor-symmetric case in (b). }
\end{figure}

 \begin {table} 
\begin{center}
    \begin{tabular}{ | c | c | c | c | c | c | c | c | c|}
    \hline
$s$ & $[0,0]$& $[1,0]$ & $[2,0]$ & $[3,0]$ &$[4,0]$&$[0,1]$&$[1,1]$&$[2,1]$ \\ \hline
     $w_{s\rightarrow s+a}^+$ & $p_0^a$ & $p_7^a$ &$p_6^a$ &$p_5^a$ &$p_4^a$& $p_3^a$&$p_2^a$&$p_1^a$\\
     \hline
     $w_{s\rightarrow s+b}^+$ & $p_0^b$ & $p_3^b$ &$p_6^b$ &$p_1^b$ &$p_4^b$& $p_7^b$&$p_2^b$&$p_5^b$ \\ \hline
    \end{tabular}
\end{center}
\caption {Zero-temperature transition rates from the state with topological charge $s$ to the states with topological charges $s+a$ and $s+b$ in the topological degeneracy. }
\label{table1}
\end {table}

\section{Tunneling current}

We now compute the tunneling current through the interferometer. We focus on the steady-state current at zero temperature and neglect the contribution from inverse tunneling processes. The tunneling current is the average of transition rates over all inequivalent states in the topological degeneracy,
\begin{equation}
\label{y9}
I=\frac{e}{4}\sum_{s,\mu} f_s w^+_{s\rightarrow s+\mu},
\end{equation}
where $s$ runs over the eight inequivalent topological charges, $\mu=a,b$, and the transition rates are given in Table~1. The probability $f_s$ that the interferometer is in the state with topological charge $s$ satisfies the master equations
\begin{equation}
\label{y10}
\frac{df_s}{dt}=\sum_{\mu} (f_{s-\mu}w^+_{s-\mu\rightarrow s}-f_{s}w^+_{s\rightarrow s+\mu}),
\end{equation}
with the normalization condition $\sum_sf_s=1$. At steady state, $df_s/dt=0$, and we solve the equations for the current. Using Fig.~2(b), we find the expression of current in the presence of flavor symmetry,
\begin{widetext}
\begin{equation}
\label{y11}
I=\frac{e}{4} R \frac{2-2c^2+\frac{c^4}{4}(1-\cos{(2\pi \Phi/\Phi_0+4\delta)})}{1-(\frac{3}{4}+\frac{1}{4\sqrt{2}})c^2+\frac{c^4}{16}[(1+\frac{1}{\sqrt{2}})(1-\cos{(2\pi \Phi/\Phi_0+4\delta)})+\frac{1}{\sqrt{2}}\sin{(2\pi \Phi/\Phi_0+4\delta)}]}.
\end{equation}
\end{widetext}
The current depends periodically on the magnetic flux through the interferometer with the period of one flux quantum $\Phi_0$. This agrees with the Byers-Yang theorem \cite{byers61}. Under the change of the sign of voltage bias at the point contacts, the current acquires an overall minus sign and the change of the sign in front of $\sin{(2\pi \Phi/\Phi_0+4\delta)}$ in the denominator. The $I$-$V$ curve is thus asymmetric, like what was found \cite{feldman06,wang10} in the Pfaffian state and in the 331 state. In Fig.~3(a), we plot the current in the 113 state with flavor symmetry and compare it with those in the Pfaffian state and in the flavor-symmetric 331 state. The current in the anti-Pfaffian state is identical to that in the Pfaffian state. \cite{wang10} We see that the current in the 113 state is more asymmetric than those in the other $\nu=5/2$ topological orders. 

\begin{figure}
\centering
\subfloat[]{\includegraphics[width=3in]{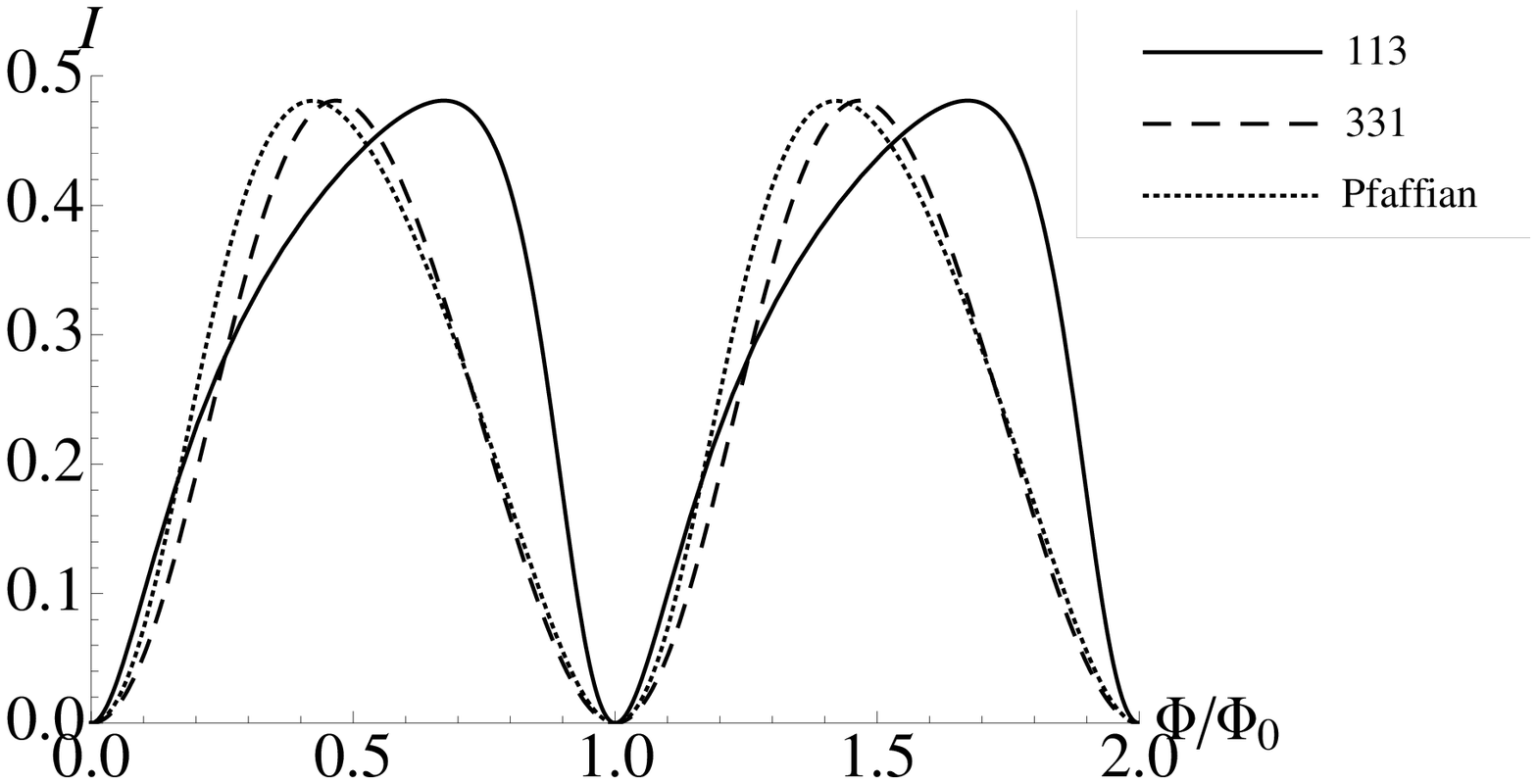}}
\quad
\subfloat[]{\includegraphics[width=3in]{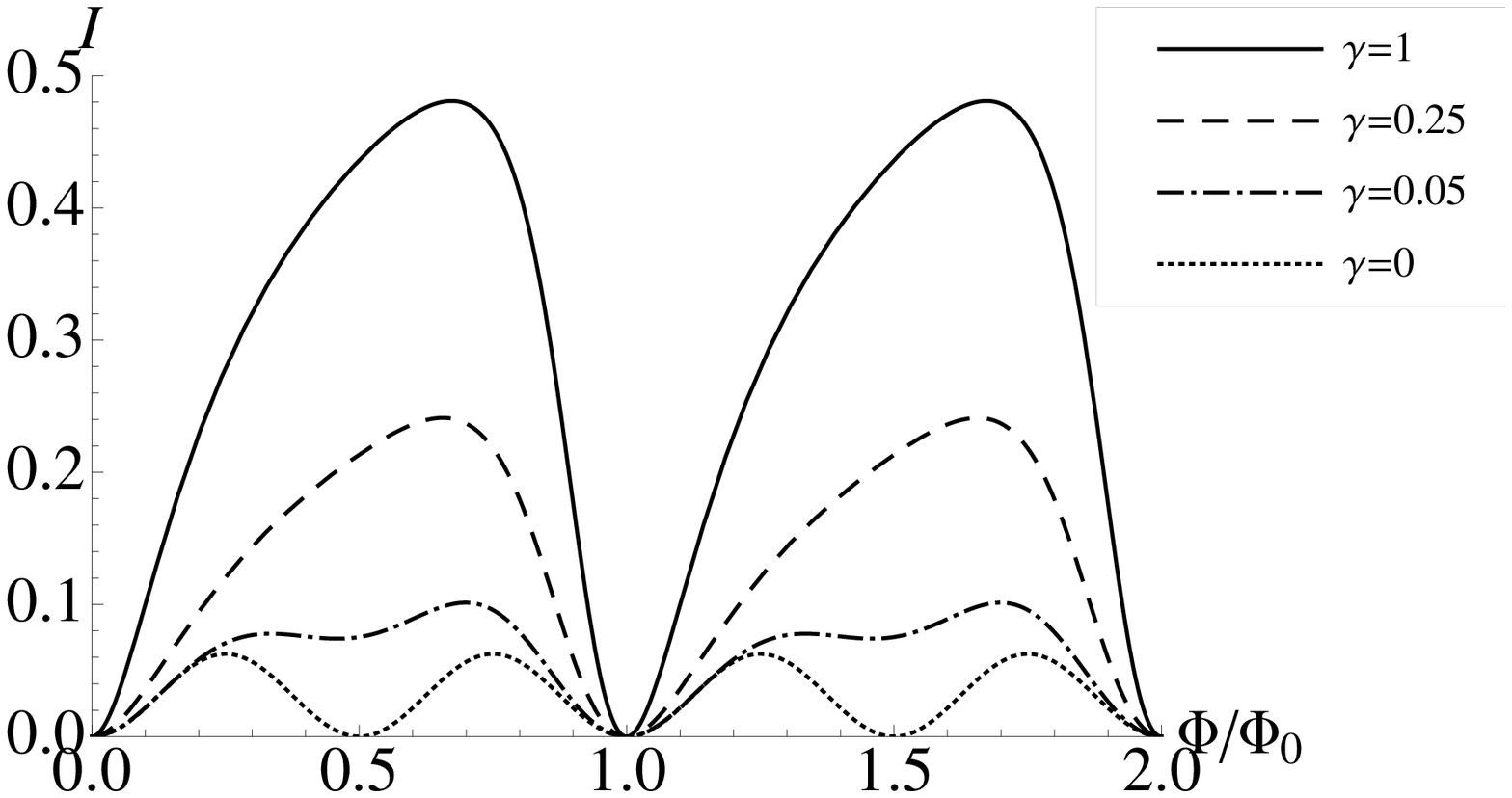}}
\caption{Steady-state tunneling current as a function of the Aharonov-Bohm flux $\Phi$. (a) Comparison of the currents in the 113 state, the 331 state, and the Pfaffian (or anti-Pfaffian) state. We assume flavor symmetry in both the 113 and 331 states. For the 113 state, we set $R=c=1$ and $\delta=0$. The current in the 331 state is acquired from Eq.~(9) in Ref.~\onlinecite{wang10}, with $u=1$ and $\delta=0$. The current in the Pfaffian state is acquired from Eq.~(8) in Ref.~\onlinecite{feldman06}, with $\lambda=1$ and $\delta=0$. For a better comparison, we have rescaled the heights of the currents in the 331 and Pfaffian states to match the current in the 113 state. (b) Current in the 113 state at different $\gamma=R^b/R^a$ values. For the cases without flavor symmetry, we set $R^a=1$, $c^a=c^b=1$, and $\delta^a=\delta^b=0$.}
\end{figure}

It is useful to quantify the asymmetry of current curves in Fig.~3(a). To this end, we notice that the currents in all three $\nu=5/2$ states can be written in the general form $I\propto\frac{1+r_1 \cos{(2\pi \Phi/\Phi_0)}}{1+r_2\cos{(2\pi \Phi/\Phi_0)}+r_3\sin{(2\pi \Phi/\Phi_0)}}=\frac{1+r_1 \cos{(2\pi \Phi/\Phi_0)}}{1+A\cos{(2\pi \Phi/\Phi_0-\phi)}}$, up to an overall factor, where $A=\sqrt{r_2^2+r_3^2}$ and $\phi=\tan^{-1}{(r_3/r_2)}$. It is easy to verify that $A<1$. The quantity $A$ characterizes the degree of asymmetry of the current curve: $A=0$ for fully symmetric current, while a large nonzero $A$ implies large asymmetry. Substituting the settings in Fig.~3(a), we find $A=0.11$, $0.28$, and $0.64$ for the 331, Pfaffian, and 113 states, respectively. 

Without flavor symmetry, the expression of tunneling current is lengthy and not enlightening. In Fig.~3(b), we plot the current at different values of $\gamma=R^b/R^a$. The minima in the current correspond to the flux at which most of the transition rates in the kinetic diagram are suppressed. The special case $\gamma=0$ is particularly interesting. In this limit, $p_k^{b}=0$ so that only quasiparticle $a$ can tunnel. The current becomes fully symmetric with the period of the Aharonov-Bohm oscillation cut in half. The new period is easily understood with the help of Fig.~2(a). When only one flavor of quasiparticles can tunnel, the system must experience eight tunneling events to complete a cycle and return to the same state at an earlier time, e.g., by following those solid arrows, with a total tunneled charge of $8\times(e/4)=2e$. This is in contrast to the situation where both flavors of quasiparticles can tunnel and a complete cycle only consists of four tunneling events. The $2e$ tunneled charge per cycle gives the $\Phi_0/2$ period of the Aharonov-Bohm oscillation, the same periodicity one finds in the physical quantities in a superconducting state with annular geometry. 

The asymmetric current at $\gamma\neq0$ and the symmetric current at $\gamma=0$ can be understood in the following. For $\gamma\neq0$, let us consider the simple limit of flavor-symmetric tunneling. Suppose now we tune the magnetic field such that $\Phi/\Phi_0$ is an integer multiple of 4; then the transition rate $p_4\approx 0$, assuming $c\approx1$ and $\delta=0$ in Eq.~(\ref{y8}). Among the other transition rates, $p_3$ and $p_5$ are relatively small compared to $p_0$, $p_1$, and $p_7$, while $p_2$ and $p_6$ are intermediate in magnitude. Imagine initially the system is in the state with topological charge $[4,0]$. After a period of time through several tunneling events, the system will return to the same initial state. More than one path in the kinetic diagram can be chosen for this return process. For example, one may follow the ``hard'' path $[4,0]\rightarrow[0,1]\rightarrow[1,1]\rightarrow[2,1]\rightarrow[4,0]$ with a smaller probability $p_4p_3p_2p_5$, or the ``easy'' path $[4,0]\rightarrow[0,1]\rightarrow[2,0]\rightarrow[2,1]\rightarrow[4,0]$ with a larger probability $p_4p_7p_6p_5$. The system can even go through multiple cycles by visiting the $[0,0]$ state before it arrives at the $[4,0]$ state for the first time. Now imagine one gradually changes $\Phi/\Phi_0$ from 0 to $1$. As $\Phi/\Phi_0$ varies, some of the easy paths deform into hard paths, and vice versa. In the 113 state, hard paths convert to easy paths at a slower rate from $\Phi/\Phi_0=0$ to 0.5 than the rate at which easy paths convert to hard paths from $\Phi/\Phi_0=0.5$ to 1. As a result, the current is asymmetric as shown in the figure. In general, the larger the inhomogeneity in the probability among different paths connecting the same initial and final states in the kinetic diagram, the larger the difference in rate between the hard-to-easy conversion of paths in the first half of the period of the Aharonov-Bohm oscillation and the easy-to-hard conversion of paths in the second half of the period, and thus the larger the asymmetry of the current. Fully symmetric $I$-$V$ curves were found in the Laughlin states, \cite{law06} where no bypaths exist in the kinetic diagrams. Our analysis finds that this is also the case in the 113 state at $\gamma=0$, which thus exhibits symmetric current.

In the 331 state or in the Pfaffian state, the probability is more balanced along different paths connecting the same two states in the kinetic diagram. Thus, the currents are less asymmetric in those topological orders than the current in the 113 state.

In practice, the shape of the current helps distinguish the 113 state from other topological orders, provided that the system is not too far away from the flavor-symmetric point. The 331 state and the Pfaffian state may not be easily distinguished via current measurement, in which case shot-noise measurement is needed, as we show in the next section. At $\gamma\approx 0$, the 113 state and the 331 state have very similar current features. Nonetheless, in this limit the Abelian orders differ from the non-Abelian orders in the periodicity of the current.

\section{Shot noise}

As shown in Ref.~\onlinecite{feldman07}, low-frequency shot noise in the Mach-Zehnder interferometer contains information about  quasiparticle statistics in the FQH state. In the following we calculate the shot noise in the 113 state at zero temperature and compare it with the results \cite{feldman07,wang10} in the Pfaffian (or anti-Pfaffian) state and in the 331 state.

We define shot noise as the Fourier transform of the current-current correlation function
\begin{equation}
\label{y12}
S(\nu)=\frac{1}{2}\int ^{+\infty}_{-\infty} \langle \hat{I}(0)\hat{I}(t)+\hat{I}(t)\hat{I}(0)\rangle e^{i \nu t}dt.
\end{equation}
The low-frequency shot noise can be related to the tunneling current through the definition of an effective charge $e^*$, $S_{\nu \rightarrow 0}=e^* I$. The ratio $e^*/e$ is the Fano factor. As we show below, the Fano factor in the 113 state can be as large as 13.6, well exceeding the maximum Fano factors in the Pfaffian (or anti-Pfaffian) state and in the 331 state.

Shot noise at low frequency can be viewed as the fluctuation in tunneled charge $Q(\tau)$ through the interferometer over a long measurement time $\tau$, 
\begin{equation}
\label{y13}
S_{\nu \rightarrow 0}=  \overline{\delta Q^2(\tau)}/\tau,
\end{equation}
where the bar denotes average over all possible tunneled charges after time $\tau$ and $\overline{\delta Q^2(\tau)}=\overline{Q^2(\tau)}-\overline{Q(\tau)}^2$. The steady-state tunneling current $I= \overline{Q(\tau)}/\tau$. Without loss of generality, let us assume that initially the topological charge at drain D2 is $[0,0]$, and that $Q(0)=0$. After time $\tau$, we may observe at D2 that $n$ quasiparticles have tunneled through the point contacts whose topological charges altogether fuse into the topological charge $s$. The tunneled electric charge during $\tau$ is then $Q(\tau)=ne/4$. Let $f_{s,n}(\tau)$ be the probability of such an observation; $f_{s,n}$ satisfies the master equations 
\begin{equation}
\label{y14}
\frac{df_{s,n}(\tau)}{d \tau}=\sum_{\mu} (f_{s-\mu,n-1}w^+_{s-\mu\rightarrow s}-f_{s,n}w^+_{s\rightarrow s+\mu}),
\end{equation}
where we note that $s$ and $n$ are not independent in $f_{s,n}$. For example, if drain D2 is found to be in the state with topological charge $s=[0,0]$ after time $\tau$, then the tunneled quasiparticles must altogether fuse into trivial topological charge and $n$ can only be an integer multiple of 4. We solve Eq.~(\ref{y14}) for the steady-state situation where $\tau$ is chosen to be long enough such that $f_{s,n}$ no longer depends on $\tau$, $df_{s,n}/d\tau=0$. Following Ref.~\onlinecite{feldman07}, we introduce the generating function $f_{s}(x,\tau)=\sum_{n} f_{s,n}(\tau)x^n$, where $n$ runs over all possible values for the given $s$. We can write
\begin{align}
\label{y15}
& \overline{Q(\tau)}= \sum_{s,n}\frac{ne}{4} f_{s,n}(\tau)=\frac{e}{4}\sum_s\left.\frac{d }{dx}f_{s}(x,\tau) \right\vert_{x=1}, \nonumber \\
&\overline{Q^2(\tau)}= \sum_{s,n}(\frac{ne}{4})^2 f_{s,n}(\tau)=(\frac{e}{4})^2 \sum_s\left. \frac{d }{dx}x\frac{d }{dx} f_{s}(x,\tau) \right\vert_{x=1},
\end{align}
and the master equations
\begin{align}
\label{y16}
\frac{df_{s}(x,\tau)}{d\tau}&=\sum_{\mu} (xf_{s-\mu}w^+_{s-\mu\rightarrow s}-f_{s}w^+_{s\rightarrow s+\mu}) \nonumber \\
& \equiv \sum_{s'}\mathbf{M}_{ss'}(x)f_{s'}(x,\tau),
\end{align}
where we have defined the kinetic matrix $\mathbf{M}(x)$, which has a finite rank. At steady state, $df_{s}(x,\tau)/d\tau=0$. Thus, $f_{s}(x,\tau)$ is the kernel of matrix $\mathbf{M}(x)$, subject to the normalization condition $\sum_sf_{s}(1,\tau)=1$. We apply the Rorbach theorem \cite{efeldman76} to solve the eigenvalue problem of $\mathbf{M}(x)$. Let $\lambda(x)$ be the largest eigenvalue of $\mathbf{M}(x)$. At $x=1$, all diagonal elements of the matrix are negative, all off-diagonal elements are non-negative, and the sum of the elements in each column equals zero. By the theorem, $\lambda(1)=0$ and is nondegenerate. All other eigenvalues are negative. At $x$ close to 1, $\lambda(x)$ is close to zero and is still nondegenerate. Thus, for large $\tau$, one can neglect the subleading terms and write $ f_{s}(x,\tau)=\eta_s \exp [\lambda(x) \tau]$, where $\eta_s$ is some constant. We find 
\begin{align}
\label{y17}
&I=\frac{e}{4}\lambda'(1), \nonumber \\
&S_{\nu \rightarrow 0}=(\frac{e}{4})^2[\lambda'(1)+\lambda''(1)], \nonumber \\
&\frac{e^*}{e}=\frac{1}{4}[1+\frac{\lambda''(1)}{\lambda'(1)}],
\end{align}
where $\lambda'(1)$ and  $\lambda''(1)$ can be obtained by differentiating the characteristic polynomial of $\mathbf{M}(x)$. \cite{feldman07}  In the presence of flavor symmetry, Eq.~(\ref{y17}) reproduces the current obtained in the previous section, and the Fano factor
\begin{equation}
\label{y18}
\frac{e^*}{e} =\frac{1}{\zeta^2}\Big[8+2\sum_{k=0}^3\frac{p_{2k+1}}{p_{2k}}
\Big(\frac{2R}{p_{2k}}+\frac{p_{2k+3}}{p_{2k+2}}\Big)-\Big(\sum_{k=0}^3\frac{p_{2k+1}}{p_{2k}}\Big)^2 \Big],
\end{equation}
where $\zeta=4+\sum_{k=0}^3(p_{2k+1}/p_{2k})$ and we have used the convention $p_k=p_{k+8l}$ with $l$ an integer. 
In Fig.~4, we plot the Fano factor in the flavor-symmetric 113 state, along with the Fano factors in the flavor-symmetric 331 state  and in the Pfaffian state. The anti-Pfaffian state has the same shot-noise profile as the Pfaffian state. We set $c=1$ for the 113 state to maximize the visibility of the Aharonov-Bohm oscillation. Experimentally, $c=1$ can be realized by adjusting the bias voltages at QPC1 and QPC2 such that $\Gamma_1=\Gamma_2$ and $\alpha=\beta$. The latter condition is fulfilled at small bias voltage $V$ and low temperature $T$, i.e., $eV,T <hv/L$, where $v$ is the velocity of the slowest edge excitation and $L$ is the difference of the distances between QPC1 and QPC2 on Edge1 and Edge2. \cite{law06,marquardt04,forster05}
In all these $\nu=5/2$ topological orders, the Fano factors are periodic functions of the magnetic flux with the period $\Phi_0$, in agreement with the Byers-Yang theorem \cite{byers61}. The Fano factor in the 113 state peaks at the height of 13.6, much higher than the Fano factor peaks at 3.2 in the Pfaffian state and 1.4 in the flavor-symmetric 331 state. The peaks of the Fano factors occur near the minima of the tunneling currents, where charge transfer is suppressed.

The large Fano factor in the 113 state arises from the same reason for the asymmetric $I$-$V$ characteristics, i.e., the large difference in the probabilities between different paths connecting the same initial and final states in the kinetic diagram, Fig.~2(b). Again, let us imagine that initially the system is in the $[4,0]$ state and $\Phi/\Phi_0$ has been tuned to be an integer multiple of 4 such that $p_4\approx 0$. In such a case, the system will dwell in the initial state for a long time before it moves to the next state via tunneling of a quasiparticle. Once the system leaves the initial state, it quickly passes through the other states in the kinetic diagram before it gets trapped again in the $[4,0]$ state for another long stay. If there were only one unique path connecting successive prolonged stays in the $[4,0]$ state and the time the system spent in the $[4,0]$ state was much longer than the total time it spent in all other states, then the effective charge $e^*$ equals the total charge tunneled in each cycle, between two successive $[4,0]$ states. This is what happens in a Laughlin state. \cite{feldman07} However, this is not the case in the 113 state with flavor symmetry. For example, there is a large probability that the system goes over multiple cycles via the states in the left half of Fig.~2(b), before it returns to the $[4,0]$ state. As a result, the effective charge $e^*$ can be very large in the 113 state. In the 331 state and the Pfaffian state, there are also bypaths connecting two states (or the same state) in the kinetic diagrams. However, the probabilities on different bypaths are close in these two topological orders, giving rise to much smaller Fano factors.

\begin{figure}
\centering
\includegraphics[width=3in]{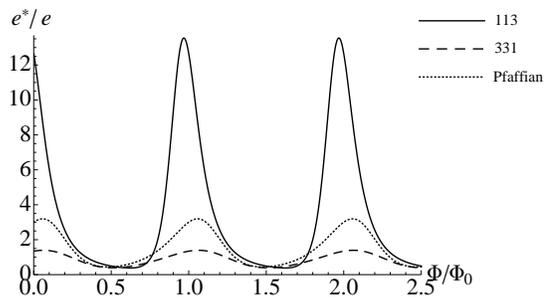}
\caption{Comparison of the Fano factors in the 113 state, the 331 state \cite{wang10}, and the Pfaffian (or anti-Pfaffian) state \cite{feldman07} as functions of the Aharonov-Bohm flux $\Phi$. We assume flavor symmetry in the 113 and 331 states.}
\end{figure}

\begin{figure}
\centering
\includegraphics[width=3in]{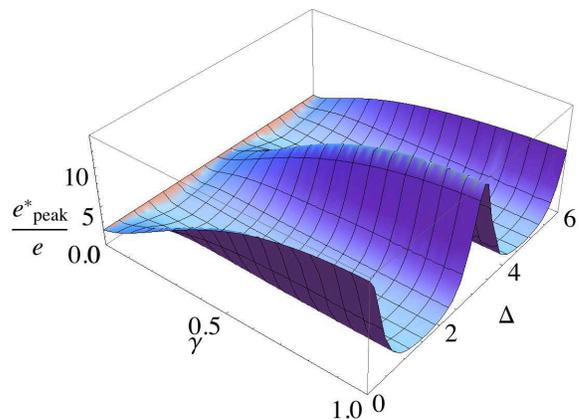}
\caption{The height $e^*_{\textrm{peak}}/e$ of the Fano factor peaks in the Aharonov-Bohm oscillation in the 113 state, as a function of $\gamma=R^b/R^a$ and $\Delta=\delta^b-\delta^a$.}
\end{figure}

In the absence of flavor symmetry, the height $e^*_{\textrm{peak}}/e$ of the Fano factor peaks in the Aharonov-Bohm oscillation is a function of the ratio $\gamma=R^b/R^a$ between tunneling amplitudes of two quasiparticle flavors and the phase difference $\Delta=\delta^b-\delta^a$, as shown in Fig.~5. $\Delta$ vanishes in the flavor-symmetric case but is nonzero in general. We find that $e^*_{\textrm{peak}}/e$ is a periodic function of $\Delta$. This is expected from Eq.~(\ref{y8}) where the phases $\delta^a,\delta^b$ are defined collinear with the Aharonov-Bohm phase. Our numerics show that the maximal value $e^*_{\textrm{max}}(\gamma)/e$ of $e^*_{\textrm{peak}}/e$ at a given $\gamma$ decreases monotonically with $\gamma$, from $e^*_{\textrm{max}}(1)/e=13.6$ to $e^*_{\textrm{max}}(0)/e=2$. At $\gamma=0$, the 113 state behaves like a Laughlin state and $e^*_{\textrm{max}}$ equals the total tunneled charge $2e$ per cycle.

In reality, neither flavor-symmetric tunneling ($\gamma=1$) nor single-flavor tunneling ($\gamma=0$) may happen. A more likely situation is in between. \cite{footnote} Nonetheless, $e^*_{\textrm{max}}/e$ well exceeds the maximum achievable Fano factor of 3.2 in the non-Abelian topological orders, provided that $\gamma>0.1$. At the same time, the domain in the parameter space for the Fano factor to exceed 2 in the 331 state is small. \cite{wang10} In general, the maximum achievable Fano factor in the 113 state is larger than those in the  331 and Pfaffian states for most values of the parameters. Experimentally, we expect such differences to be measurable with current instrumental precision \cite{heiblum09}.

\section{Conclusions}

In conclusion, we have shown that an electronic Mach-Zehnder interferometer can be used as a tool to identify different topological orders at $\nu=5/2$. We have calculated the zero-temperature tunneling current and shot noise through the interferometer in the Halperin 113 state and compared the results with those in the Halperin 331 state and in the non-Abelian Pfaffian and anti-Pfaffian states, the latter two states having identical interference characteristics. We find that the $I$-$V$ curve in the 113 state is more asymmetric than those in the 331 state and in the Pfaffian state. In addition, the maximum Fano factor of 13.6 in the 113 state, found in the case of flavor-symmetric quasiparticle tunneling, is much greater than the maximum Fano factors 2.3 in the 331 state and 3.2 in the Pfaffian and anti-Pfaffian states. In practice, the combination of tunneling current and shot-noise measurements can provide clear discrimination of these $\nu=5/2$ topological orders.

\begin{acknowledgments}

We gratefully thank D. E. Feldman for encouragement and helpful discussions on this project. This work was partly supported by the NSF under Grant No. DMR-1205715.

\end{acknowledgments}


\begin{thebibliography}{99}
\bibitem{footnote} If the quasiparticle flavors depend on electron spin species then $\gamma$ may be adjusted by controlling the spin polarization of the current.
\end{thebibliography}

\end{document}